\documentstyle[12pt]{article}

\newcommand{\ds}{\displaystyle}
\newcommand{\ltwid}{\raise.3ex\hbox{$<$\kern-.75em\lower1ex\hbox{$\sim$}}}
\newcommand{\rtwid}{\raise.3ex\hbox{$>$\kern-.75em\lower1ex\hbox{$\sim$}}}

\title{Persistence of pseudogap formation
       in quasi-2D systems with arbitrary carrier density}
\author{{\sl Rachel~M.~Quick
        and Sergei~G.~Sharapov$^\dagger$}\\
{\sl Department of Physics, }\\
{\sl University of Pretoria,}\\
{\sl 0002 Pretoria, South Africa}}

\date{December 10, 1997}

\begin{document}

\maketitle


\begin{abstract}
The existence of a pseudogap above the critical temperature has been
widely used to explain the anomalous behaviour of the normal state of
high-temperature superconductors. In two dimensions the existence of a
pseudogap phase has already been demonstrated in a simple model.
It can now be shown that the pseudogap phase persists even
for the more realistic case where coherent interlayer tunneling is taken
into account.  The effective anisotropy is surprisingly large and even
increases with increasing carrier density.
\end{abstract}



{\sl* Corresponding author: R.M.~Quick \\
Department of Physics, University of Pretoria,\\
Pretoria 0002 SOUTH AFRICA \\
E-mail: rcarter@scientia.up.ac.za}

\eject


\section{Introduction}

The anomalous behaviour of the normal state of high-temperature
superconductors (HTSC) \cite{Loktev.review,Pines.review}
(including the behaviour of the spin susceptibility, resistivity,
specific heat and photo-emission spectra) has been recently interpreted
in terms of the formation of a pseudogap above the critical temperature,
$T_c$ \cite{Levi,Randeria.Nature}.

The formation of a pseudogap phase above $T_c$ has been explicitly
demonstrated in a model non-relativistic 2D fermi-system
\cite{Gusynin,LSh.review}. The work is based on the peculiarities of
the Berezinskii-Kosterlitz-Thouless (BKT) phase formation (see also
\cite{MacKenzie}), which is a two stage process. For a 2D system
one must rewrite the order parameter $\Phi(x)$, where
$x={\bf r},\tau$ denotes the position
and imaginary time, in terms of its modulus $\rho(x)$ and its phase
$\theta(x)$ i.e. $\Phi(x) = \rho(x) \exp[- i \theta(x)]$. This was originally
stated by Witten in the context of 2D quantum field theory \cite{Witten}.
It is clearly impossible to obtain
$\Phi \equiv \langle \Phi(x) \rangle \ne 0$
at finite $T$ since this corresponds to the formation of homogeneous
long-range (superconducting) order which is forbidden by the
Coleman-Mermin-Wagner-Hohenberg theorem.  However it is possible to
obtain $\rho \equiv \langle \rho(x) \rangle \ne 0$ but at the same
time $\Phi = \rho \langle \exp[-i \theta(x)] \rangle = 0$
due to random fluctuations in the phase $\theta(x)$. We stress that
$\rho \ne 0$ does not imply long-range superconducting order
(which is destroyed by the phase fluctuations) and is therefore
not in contradiction with the above-mentioned theorem.

Thus one has three regions in the 2D phase diagram \cite{Gusynin}. The
first is the superconducting (BKT) phase, where $\rho \ne 0$. In this region
there is algebraic order and a power law decay of the correlations. The second
is the pseudogap phase where $\rho$ is still non-zero but the correlations
decay exponentially. The third is the normal (Fermi-liquid) phase where
$\rho = 0$. Note that $\Phi = 0$ everywhere. The unusual properties of
the second region, which lies between the superconducting and normal
phases, have previously been used in \cite{Gusynin} to explain the pseudogap
behaviour in HTSC. For example in the mean-field calculation of the
paramagnetic susceptibility \cite{LSh.review} the parameter $\rho$ plays
the role of the energy gap $\Delta$ in the theory of ordinary superconductors.
Thus in this calculation one has the opening of an energy gap $\rho$ (or
equivalently a lowering of the density of states) above the superconducting
transition temperature.

The description of the phase fluctuations and the BKT transition closely
resembles that given by Emery and Kivelson \cite{Emery}. However in their
phenomenological approach the field $\rho(x)$ does not appear while in the
present microscopic approach based on \cite{Witten} it appears rather
naturally.

Although the description in terms of modulus and phase variables is essential
from a mathematical point of view, the physical implications for the theory
of superconductivity have not been fully understood.  Previous work
\cite{Witten} was at zero temperature but the application to
condensed matter requires the extension to finite temperature where
$\rho$ is a function of the temperature $T$.  It makes sense therefore
to define the temperature $T_\rho$ at which $\rho$ becomes zero. This
temperature is then interpreted as the temperature at which the pseudogap
opens. Since in \cite{Gusynin,MacKenzie} and in the present work $\rho(x)$
has only been treated in the mean-field approximation i.e. one has neglected
the fluctuations in both $\rho(x)$ and $\theta(x)$, a second-order
phase transition was obtained. However it is well known experimentally that
the formation of the pseudogap phase does not display any sharp transition. It
can be argued however that the fluctuations may convert the obtained sharp
transition to a crossover \cite{Gusynin}.

An important question is whether the pseudogap phase
(PP) forms in real HTSC, which are only quasi-2D systems. Another related
question is whether the pseudogap phase, if present, remains
large enough to explain the experimentally observed anomalies even when
interlayer tunneling is included.

This paper will study  PP formation for a quasi-2D system with
arbitrary carrier density. Note that this problem is further complicated
by the formation in quasi-2D systems of the ordinary bulk
superconducting phase with homogeneous long-range order (LRO) at critical
temperature $T_c$ (see for example \cite{Ichinose}). Since the BKT phase
formation corresponds to the formation of two-dimensional order while true
LRO is three dimensional one expects that for weakly coupled layers
$T_c \le T_{\rm BKT}^q$, where $T_{\rm BKT}^q$ denotes the
critical temperature for the BKT transition in the quasi-2D system.
$T_{\rm BKT}^q$ represents the temperature at which the system becomes
superconducting in the layers (planes) while $T_c$ is the critical temperature
for bulk superconductivity. Thus $T_{\rm BKT}^q$ is the maximum temperature
at which superconducting behaviour is present. The temperature range
$T_c \le T \le  T_{\rm BKT}^q$ corresponds to a crossover region where the
effect of the interlayer tunneling is  insufficient to produce three
dimensional behaviour and the behaviour remains
BKT-like. There is in fact some experimental evidence for the existence of
this crossover region ($\sim 1$K) in optimally doped
YBa$_2$Cu$_3$O$_{7-\delta}$ \cite{Stamp}.

The size of this region is obviously critically dependent on the
anisotropy, the carrier density and the precise form of the interlayer
tunneling. At very low densities (i.e. in the Bose limit of isolated pairs)
and for physically reasonable values for the parameters, it has been shown for
quasi-2D system \cite{LSh.review,GLSh.PhysicaC,Turkowski} that long-range
order is only established well below $T_{\rm BKT}$. However these densities
are not realised physically. It can also be shown that $T_c$ approaches the
BCS critical temperature asymptotically in the high-density limit
\cite{Kats} so that both this region and the pseudogap phase must vanish
asymptotically. At the intermediate carrier densities found in HTSC we find
that the pseudogap phase remains large enough to explain the  observed
anomalies. However the experimentally observed superconducting transition is
practically three dimensional \cite{Stamp} so that one expects
$T_c \ltwid T_{\rm BKT}^q$ at these densities and this is the subject of
current investigations.

For these reasons we only calculate the temperatures $T_{\rm BKT}^q$ and
$T_\rho$ as functions of the carrier density $n_f$ to establish the boundaries
of the PP. Here $T_{\rm BKT}^q$ denotes the critical temperature for the BKT
transition in the quasi-2D system.

We show that the value of $T_{\rho}$ is practically identical in the 2D and
the quasi-2D  systems, while for realistic model parameters
$T_{\rm BKT}^q > T_{\rm BKT}$. More importantly, even at relatively low
anisotropy (for example that observed in the compound
YBa$_{2}$Cu$_{3}$O$_{7-\delta}$),
the difference between $T_{\rm BKT}^q$ and $T_{\rm BKT}$ is too small
to destroy the PP at all carrier densities. In addition the difference at
high doping levels can be shown to tend to zero logarithmically. This
enables one to seriously consider the pseudogap phase explanation claimed in
reference \cite{Gusynin} even in the quasi-2D case.

\section{Model and Formalism}

The nature of the interplane tunneling in HTSC is not yet well established
\cite{Marel1} and several different models exist. Here we choose the
simplest possible Hamiltonian density, often employed for studying
HTSC \cite{GLSh.PhysicaC,Varlamov},
\begin{equation}
H = -\psi_{\sigma}^{\dagger}(x)
\left[ \frac{\nabla_{\perp}^{2}}{2 m_{\perp}} +
\frac{1}{m_{z} d^{2}} \cos(i d \nabla_{z}) + \mu \right] \psi_{\sigma}(x)
- V \psi_{\uparrow}^{\dagger}(x) \psi_{\downarrow}^{\dagger}(x)
    \psi_{\downarrow}(x) \psi_{\uparrow}(x),     \label{Hamiltonian}
\end{equation}
where $x \equiv  \tau, \mbox{\bf r}_{\perp}, r_{z}$
(with $\mbox{\bf r}_{\perp}$ being a 2D vector);
$\psi_{\sigma}(x)$ is a fermion field,
$\sigma = \uparrow, \downarrow$ is the spin variable;
$m_{\perp}$ is the effective carrier mass in the planes
(for example CuO$_{2}$ planes);
$m_{z}$ is an effective mass in the $z$-direction;
$d$ is the interlayer distance;
$V$ is an effective local attraction constant;
$\mu$ is a chemical potential which fixes the carrier density $n_{f}$;
and we take $\hbar = k_{B} = 1$.

The Hamiltonian proposed proves to be very convenient for the study of
fluctuation stabilization by weak 3D one-particle inter-plane tunneling.
We have omitted in (\ref{Hamiltonian}) the two particle (Josephson)
tunneling considering it to be less important than the one-particle
coherent tunneling already included. There can be situations where
Josephson tunneling is more important. In fact some authors consider
the most important mechanism for HTSC to be the incoherent inter-plane
hopping (through, for instance, the impurity (localized) states or due
to the assistance of phonons). We will not however consider Josephson
tunneling here. We do however take into account the layered structure
of HTSC which is a vitally important extension to the 2D models usually
considered.

It is significant that the large anisotropy in the conductivity cannot
be identified with the corresponding anisotropy in the effective masses
$m_{z}$ and $m_{\perp}$. In particular, HTSC with rather large anisotropy
in the $z$-direction do not display the usual metal behaviour at low
temperatures \cite{Watanabe}. However this semiconducting behaviour
is not directly related to the pseudogap phenomena \cite{Watanabe} and
the Hamiltonian (\ref{Hamiltonian}) may be used to study the qualitative
features of pseudogap opening.

The Hubbard-Stratonovich method was applied to study the system described
by (\ref{Hamiltonian}). In this method the statistical sum $Z({v}, \mu, T)$
is given as a functional integral over the Fermi-fields (Nambu spinors) and
the auxiliary field $\Phi(x) = V \psi_{\uparrow}^{\dagger}
\psi_{\downarrow}^{\dagger}$.  In contrast to the usual method where one
calculates $Z$ in terms of the $\Phi(x)$ and $\Phi^{\ast}(x)$ variables, the
parameterisation $\Phi(x) = \rho(x) \exp{[- i \theta(x)]}$ should
be used \cite{Witten} (see also \cite{Thouless,Beck}).  In addition to this
reparameterisation one must make the replacement
$\psi_{\sigma}(x) = \chi_{\sigma}(x) \exp{[i \theta(x)/2]}$. This
representation splits the charged fermion field $\psi_{\sigma}(x)$ into a
neutral fermion field $\chi_\sigma(x)$ and a charged boson field
part $\exp{[i \theta(x)/2]}$. This resembles the spinons and holons
in Anderson's approach.

This particular choice of parameterisation ensures that $\Phi(x)$ is
single-valued with period $2 \pi$. As a result one obtains
\begin{equation}
Z(v, \mu, T) = \int \rho {\cal D} \rho {\cal D} \theta
\exp{[-\beta \Omega (v, \mu, T, \rho(x), \partial \theta (x))]},
                                         \label{statsum.pase}
\end{equation}
where
\begin{equation}
\Omega  =  \frac{T}{V} \int_{0}^{\beta}
d \tau \int d \mbox{\bf r} \rho^{2} - T \mbox{Tr Ln} G^{-1}
                        \label{Effective.Action}
\end{equation}
is the one-loop effective action which now depends on the modulus-phase
variables. The action (\ref{Effective.Action}) is expressed in terms of
the Green function $G$ of the initial (charged) fermions which now has
the following operator form $G^{-1} = {\cal G}^{-1} - \Sigma$ where
\begin{eqnarray}
& & {\cal G}^{-1}[\rho(x)] = - \hat{I} \partial_{\tau} + \tau_{3}
\left[\frac{\nabla_{\perp}^{2}}{2 m_{\perp}} +
\frac{1}{m_{z} d^{2}} \cos(i d \nabla_{z}) + \mu \right] +
\tau_{1} \rho(x); \label{neutg} \\
& & \Sigma[\partial \theta(x)] =
\tau_{3} \left[\frac{i \partial_{\tau} \theta}{2} +
\frac{(\nabla_{\perp} \theta)^{2}}{8 m_{\perp}} +
\frac{(\nabla_{z} \theta)^{2}}{8 m_{z}}
\cos(i d \nabla_{z}) \right] - \nonumber \\
& & \hat{I} \left[
\frac{i \nabla_{\perp}^{2} \theta}{4 m_{\perp}} +
\frac{i \nabla_{z}^{2} \theta}{4 m_{z}} \cos(i d \nabla_{z}) +
\frac{i \nabla_{\perp} \theta  \cdot \nabla_{\perp}}{2 m_{\perp}} +
\frac{i \nabla_{z} \theta \sin(i d \nabla_{z})}{2 m_{z} d}
\right].                       \label{Sigma}
\end{eqnarray}
Here $\cal G$ is the Green function for the neutral fermions.

Note that in $\Sigma$ we have omitted higher order terms in
$\nabla_{z} \theta$ but in order to keep all relevant terms in the
expansion of $\sin(i d \nabla_{z})$ the necessary resummation was done.
Since the low-energy dynamics in the phases in which
$\rho \neq 0$ is determined by the long-wavelength fluctuations
of $\theta(x)$, only the lowest order derivatives of the phase need be
retained in what follows.  This gives the one-loop effective action as
\begin{equation}
\Omega \simeq
\Omega _{kin} (v, \mu, T, \rho, \partial \theta) +
\Omega _{pot}^{\mbox{\tiny MF}} (v, \mu, T, \rho)
                  \label{kinetic.phase+potential}
\end{equation}
where
\begin{equation}
\Omega _{kin} (v, \mu, T, \rho, \partial \theta)
 =  T \mbox{Tr} \sum_{n=1}^{\infty}
\left. \frac{1}{n} ({\cal G} \Sigma)^{n}
\right|_{\rho = \: \mbox{\tiny const}}
               \label{Omega.Kinetic.phase}
\end{equation}
and
\begin{equation}
\Omega _{pot}^{\mbox{\tiny MF}} (v, \mu, T, \rho)  =
\left. \left(\frac{1}{V} \int d \mbox{\bf r} \rho^{2} -
T \mbox{Tr Ln} {\cal G}^{-1} \right) \right|_{\rho = \: \mbox{\tiny const}}.
                 \label{Omega.Potential.modulus}
\end{equation}

Given the representation (\ref{kinetic.phase+potential}) one can obtain
the full set of equations for $T_{\rm BKT}^q$,
$\rho(T_{\rm BKT}^q)$ and $\mu(T_{\rm BKT}^q)$ at given
$\epsilon_{F}$. While the equation for $T_{\rm BKT}^q$
only depends on the kinetic part (\ref{Omega.Kinetic.phase}) of the
effective action, the equations for $\rho(T_{\rm BKT}^q)$ and
$\mu(T_{\rm BKT}^q)$ can to a good approximation be obtained using
the mean field potential (\ref{Omega.Potential.modulus}).
In the phase where $\rho \neq 0$ the mean-field approximation for $\rho$
describes the system well due to the nonperturbative character of the
Hubbard-Stratonovich method.

We note that the expression for the potential (\ref{Omega.Potential.modulus})
in terms of $\rho^2$ is identical to the mean-field potential in the
BCS approximation but with $|\Phi|^2$ replaced by $\rho^2$ \cite{LSh.review}.
Thus $T_\rho$, the temperature at which $\rho = 0$, is in this approximation
identical to the BCS mean field temperature $T^{MF}_c$. However, although
$T_\rho = T^{MF}_c$ in the mean-field approximation for $\rho(x)$,
the two temperatures have a very different basis, both
mathematically and physically. This becomes evident when one includes the
fluctuations. Not only does $T_\rho$ remain finite in two dimensions
due to the structure of the perturbation theory in the new modulus-phase
variables, but it is also bounded below by $T_{\rm BKT}$. On the other hand
$T^c_{MF}$ approximates the temperature $T_c$ where $\Phi$ becomes non-zero
(onset of long-range order) which is zero in two dimensions.

The modulus-phase representation introduced here
is a good tool to consider different types of short range order.
The success of the BCS approximation is related to the fact that in 3D system
with large carrier density short and long range order set in
simultaneously because the fluctuations do not change the situation
drastically.

\section{The Berezinskii-Kosterlitz-Thouless transition in quasi-2D theory}

If the model under consideration reduced to some known model
describing the BKT phase transition, we could easily write the equation
for $T_{\rm BKT}$. Indeed, in the lowest orders the kinetic term
(\ref{Omega.Kinetic.phase}) coincides with classical spin
quasi-2D XY-model \cite{Hikami} (see also \cite{Ichinose}) which has
the following continuum Hamiltonian
\begin{equation}
H = \frac{1}{2d} \int d \mbox{\bf r}
\{ J [\nabla_{\perp} \theta(\mbox{\bf r})]^{2} +
J_{z} [\nabla_{z} \theta(\mbox{\bf r})]^{2} \}.
                             \label{XY.Hamiltonian}
\end{equation}
Here $J$ and $J_{z}$ are constant coefficients ($J_{z} \ll J$).
Unfortunately the temperature for the BKT transition in the quasi-2D
case is not as well investigated as in the pure 2D case. In fact only
in the highly anisotropic case, $\alpha \equiv J_{z}/J \ll 1$, when
the vortex ring excitations are irrelevant has the transition temperature
been derived \cite{Hikami}. In this limit the temperature $T_{\rm BKT}^{q}$
for the BKT transition in the quasi-2D system is close to that in the pure 2D
case and determined by the equation
\begin{equation}
T_{\rm BKT}^{q} = \frac{\pi}{2} J
\left[1 + \frac{8 \pi}{\ln^{2} \alpha} \right].
                                 \label{BKT.quasi}
\end{equation}
This equation was given in \cite{Hikami} and employed to calculate $T_{\rm
BKT}^q$ for the relativistic quasi-2D four-Fermi theory \cite{Ichinose}.
The equation was derived using the renormalization
group technique, which takes into account the non-single-valuedness of the
phase $\theta$.
Thus, the fluctuations of the phase are taken into account at a higher
approximation than Gaussian.
Below the temperature $T_{\rm BKT}^{q}$ the correlation function
\begin{equation}
\langle e^{i \theta(\mbox{\bf r})}  e^{-i \theta(0)} \rangle \to
\alpha^{T/4 \pi J} \qquad r \to \infty,
                                \label{correlator}
\end{equation}
while above $T_{\rm BKT}^{q}$ this correlator decreases exponentially.
This is the BKT transition in the classical quasi-2D XY model.
The small correction to the unit in the brackets of
Eq. (\ref{BKT.quasi}) corresponds to
the influence of the third direction. We note that the equation is
only correct in the limit $\alpha \ll 1$. From a physical point of
view it is evident that the BKT phase cannot form for $\alpha \sim 1$, which
corresponds to the 3D limit. Moreover, if the temperature $T_c$ of the LRO
formation approaches $T_{\rm BKT}^q$ from below, the BKT phase will not
form and one will have a superconducting transition directly into a phase
with LRO. This may well correspond to the experimental situation and we hope
to study this question in detail in our future work.

To expand $\Omega _{kin}$ up to $\sim (\nabla \theta)^{2}$, it is
sufficient to consider only the terms with $n=1,2$ in the expansion
(\ref{Omega.Kinetic.phase}). The method is the same as that
in \cite{Gusynin,Schakel}, and gives
\begin{eqnarray}
\Omega _{kin} = \frac{T}{2 d} \int_{0}^{\beta} d \tau \int d \mbox{\bf r}
&& \left[ J(\mu, T, \rho(\mu, T)) (\nabla_{\perp} \theta)^{2} +
J_{z}(\mu, T, \rho(\mu, T)) (\nabla_{z} \theta)^{2} + \right.
\nonumber                       \\
&& \left. K(\mu, T, \rho(\mu, T)) (\partial_{\tau} \theta)^{2}
+ n_{F}(\mu, T, \rho(\mu, T)) i \partial_{\tau} \theta \right],
                 \label{Omega.Kinetic.phase.final}
\end{eqnarray}
where
\begin{equation}
J(\mu, T, \rho) = \frac{d}{4m_{\perp}} n_{F}(\mu, T, \rho) -
\frac{T}{8 \pi^{2}} \int_{0}^{2 \pi} d t
\int_{- \frac{\ds \mu}{\ds 2T} -
\frac{\ds w \cos t}{\ds 2 T}}^{\infty} dx
\frac{x + \mu / 2T + w \cos t/(2 T)}
{\cosh^{2} \sqrt{x^{2} + \frac{\ds \rho^{2}}{\ds 4 T^{2}}}},
                                              \label{J}
\end{equation}
\begin{eqnarray}
J_{z}(\mu, T, \rho) & = &
\frac{m_{\perp}}{4 m_z} \frac{1}{(2 \pi)^{2}} \left[ \int_{0}^{2 \pi} dt
\cos t \left\{w \cos t +
\sqrt{\left(w \cos t + \mu
\right)^{2} + \rho^{2}} \right. \right.
\nonumber               \\
& + & \left. 2 T \ln{\left[1 + \exp{\left(-\frac{\sqrt{(w
\cos t +\mu)^{2} + \rho^{2}}}{T}\right)} \right]} \right \}
\nonumber                \\
& - & \left. w \int_{0}^{2 \pi} d t \sin^2 t
\int_{- \frac{\ds \mu}{\ds 2T} -
\frac{\ds w \cos t}{\ds 2 T}}^{\infty} dx
\frac{1} {\cosh^{2} \sqrt{x^{2} + \frac{\ds \rho^{2}}{\ds 4 T^{2}}}} \right],
                                      \label{Jz}
\end{eqnarray}
\begin{equation}
K(\mu, T, \rho) = \frac{m_{\perp}}{(4 \pi)^{2}} \int_{0}^{2 \pi} d t
\left(1 +  \frac{w \cos t + \mu}{\sqrt{(w \cos t + \mu)^{2} + \rho^{2}}}
\tanh \frac{\sqrt{(w \cos t + \mu)^{2} + \rho^{2}}}{2 T}
\right)
                        \label{K}
\end{equation}
and where we have introduced the bandwidth in the $z$-direction
$w = (m_z d^2)^{-1}$.
Here
\begin{eqnarray}
n_{F}(\mu, T, \rho) & = &
\frac{m_{\perp}}{(2 \pi)^{2} d} \int_{0}^{2 \pi} dt
\left\{\mu + \sqrt{\left(w \cos t + \mu \right)^{2} +
\rho^{2}} \right.
\nonumber                   \\
& + & \left. 2 T \ln{\left[1 + \exp{\left(-\frac{\sqrt{(w
\cos t +\mu)^{2} + \rho^{2}}}{T}\right)} \right]}\right\}
                              \label{fermion.density}
\end{eqnarray}
takes the form of a fermi-quasiparticle density. Note that
$J(\mu,T, \rho=0)=J_{z}(\mu,T,\rho=0)=0$.
This property of the phase stiffness is present in the 2D
model where it implies that $\rho \ne 0$ at $T_{\rm BKT}$ and thus $T_\rho
\ge T_{\rm BKT}$. The continued presence of this property
in the quasi-2D case implies that the essentially nontrivial cosine
dispersion law for motion in the third direction is correctly treated.

Now we discuss the features of the BKT transition in our model.
In contrast to the Hamiltonian (\ref{XY.Hamiltonian}) the field
$\theta$ is also depend on the imaginary time $\tau$, and therefore
by Fourier decomposition, one can write
\begin{equation}
\theta(\tau, \mbox{\bf r}) = \sum_{n = - \infty}^{\infty}
\exp(i 2 \pi n T \tau) \theta_{n} (\mbox{\bf r}.)
                                \label{theta.Fourier}
\end{equation}
From Eq. (\ref{Omega.Kinetic.phase.final}) we can see that the nonzero
mode $\theta_{n}(\mbox{\bf r})$ $n \neq 0$ has a mass
$m_{n}^{2} \sim (2 \pi n T)^{2} K$, and only the massless component
$\theta_{0} (\mbox{\bf r})$ is relevant in the low-energy
region. In terms of the $\theta_{n}$ fields, the effective {\em classical}
action is expressed by
\begin{eqnarray}
\Omega _{kin} = \frac{1}{2 d} \int d \mbox{\bf r} \!\!\!\!
&& \left[ J (\nabla_{\perp} \theta_{0})^{2} +
J_{z} (\nabla_{z} \theta_{0})^{2} + \right.
\nonumber                       \\
&& \left. \sum_{n \neq 0} \left\{
J (\nabla_{\perp} \theta_{-n})(\nabla_{\perp} \theta_{n}) +
J_{z} (\nabla_{z} \theta_{-n})(\nabla_{z} \theta_{n}) +
m_{n}^{2} \theta_{-n} \theta_{n} \right\} \right].
                   \label{Omega.Kinetic.phase.n}
\end{eqnarray}
Recalling that $\theta$ is an angular variable, we see that the
$\theta_{0}$-part of the effective classical action
(\ref{Omega.Kinetic.phase.n}) is nothing but
the Hamiltonian of the quasi-2D XY model (\ref{XY.Hamiltonian}).
This makes it possible to write the equation for $T_{\rm BKT}^q$:
\begin{equation}
T_{\rm BKT}^q=\frac{\pi}{2} J(\mu, T_{\rm BKT}^q,\rho)
\left[1 + \frac{8 \pi}{\ln^{2} \alpha(\mu,T_{\rm BKT}^q,\rho)}
\right],                      \label{BKT.equation}
\end{equation}
where $\alpha(\mu,T,\rho) = J_{z}(\mu,T,\rho)/J(\mu,T,\rho)$
is a function of $\mu$, $T$ and $\rho(\mu, T)$.
Recall that equation (\ref{BKT.equation}) is only correct in the
limit $\alpha \ll 1$. Thus after the calculation of $T_{\rm BKT}^q$ one
must check that the condition $\alpha \ll 1$ is satisfied.

$\theta_{0}$ has non-trivial dynamics described by
the correlation function
\begin{equation}
\langle e^{i \theta_0(\mbox{\bf r})}  e^{-i \theta_0(0)} \rangle \to
\alpha^{T/4 \pi J} \qquad r \to \infty,
                                \label{correlator2}
\end{equation}
which is identical to (\ref{correlator}) but with $\theta$ replaced by
$\theta_0$. However the gauge-invariant
correlation function, $\langle \Phi^{\ast}(\tau =0, \mbox{\bf r})
\Phi(0) \rangle$, has an additional factor from the massive modes $\theta_n,
n \neq 0$ and, regardless of the correlator for $\theta_0$ (\ref{correlator2}),
one has
\begin{equation}
\langle \Phi^{\ast}(\tau =0, \mbox{\bf r}) \Phi(0) \rangle \to 0
\qquad r \to \infty
                          \label{correlator.Phi}
\end{equation}
for any $T>T_c$ (see the details in \cite{Ichinose}) which is consistent
with the assumption that LRO is absent.

Although mathematically the problem reduces to a known problem, the analogy
is incomplete. Indeed, in the XY model the vector (spin) subject to ordering
is assumed to be a unit vector with no dependence on $T$. In our model
this is not the case, and a self-consistent calculation of $T_{\rm BKT}$ as
a function of $n_{f}$ requires additional equations for $\rho$ and $\mu$,
which together with (\ref{BKT.equation}) form a complete system.

Using the definition (\ref{Omega.Potential.modulus}), one
obtains (see e.g. \cite{Gusynin})
\begin{equation}
\Omega _{pot}(v, \mu, T, \rho) = {v} \left[
\frac{\rho^2}{V} -
\int \frac{d \mbox{\bf k}}{(2 \pi)^{3}}
\left\{2T
\ln\cosh{\frac{\sqrt{\xi^{2}(\mbox{\bf k})     + \rho^{2}}}{2 T}} -
\xi(\mbox{\bf k}) \right\} \right],
                     \label{Omega.Potential.modulus.final}
\end{equation}
where $\xi(\mbox{\bf k}) = \varepsilon(\mbox{\bf k}) - \mu$ with
$\varepsilon (\mbox{\bf k}) = \mbox{\bf k}_{\perp}^{2}/2m_{\perp}
- w \cos k_{z} d$.
Then the missing equations are
the condition $\partial \Omega_{pot}(\rho)/ \partial \rho = 0$
and the equality $v^{-1} \partial \Omega_{pot} /\partial \mu = -n_f$, which
fixes $n_f$ :
\begin{equation}
\frac{1}{V} = \int \frac{d \mbox{\bf k}}{(2 \pi)^{3}}
\frac{1}{2\sqrt{\xi^{2}(\mbox{\bf k}) + \rho^{2}}}
\tanh{\frac{\sqrt{\xi^{2}(\mbox{\bf k}) +
\rho^{2}}}{2 T}},              \label{rho}
\end{equation}
\begin{equation}
n_{F}(\mu, T, \rho) = n_f.     \label{number.rho}
\end{equation}
The equations (\ref{rho}) and (\ref{number.rho}) obtained above
comprise a self-consistent system for determining the modulus $\rho$
of the order parameter and the chemical potential $\mu$ in the
mean-field approximation for fixed $T$ and $n_{f}$.
In the phase where $\rho \neq 0$ the mean-field approximation for $\rho$
describes the system well due to the nonperturbative character of the
Hubbard-Stratonovich method and the character of the perturbation theory in
the modulus-phase variables.

To simplify the problem one can take the limit $m_{z} \to \infty$ in the
expressions (\ref{J}), (\ref{fermion.density}) assuming that
$m_{z}/m_{\perp} \gg 1$ and $w = m_{z}^{-1} d^{-2} \ll T_{\rm BKT}^q$. This is
indeed the case in real HTSC compounds e.g. for $m_{z} \approx 10^{2} m_{e}$
and $d=10 \dot A$ the value of $\hbar^{2}/(m_{z}d^{2}k_{B}) \sim 10$K is far
less than the critical temperature. This simplification is useful but not
essential since the numerical results are practically unchanged by this
approximation.

The energy of two-particle bound states in vacuum
$\varepsilon_{b} = -2W \exp (-4 \pi d/m_{\perp} V)$
(see e.g. \cite{GGL}) where $W$ is the bandwidth
in the plane, is more convenient than the four-fermi constant $V$.
For example, one can easily take the limits $W \to \infty$ and $V \to 0$
in the equation (\ref{rho}), which after this renormalization becomes
(in the limit $m_{z} \to \infty$)
\begin{equation}
\ln{\frac{|\varepsilon_{b}|}{\sqrt{\mu^{2} + \rho^{2}} - \mu }} =
2 \int_{-\mu/T}^{\infty} d u
\frac{1}{\ds \sqrt{u^2 + \left(\frac{\ds \rho}{\ds T}\right)^2}
\left[\exp{\sqrt{u^2 + \left(\frac{\ds \rho}{\ds T}\right)^2}} + 1 \right]}.
                      \label{rho.bound}
\end{equation}
Thus, in practice, we will solve numerically the system of equations
(\ref{BKT.equation}), (\ref{rho.bound}) and (\ref{number.rho}) to study
$T_{\rm BKT}^q$ as a function of $n_{f}$.

Setting $\rho = 0$ in the equations (\ref{rho}) and
(\ref{number.rho}), we arrive (in the same approximation) at the
equations for the critical temperature $T_{\rho}$ above which $\rho =0$
and the corresponding value of $\mu$:
\begin{equation}
\ln{\frac{|\varepsilon_{b}|}{T_{\rho}}\frac{\gamma}{\pi}} =
- \int_{0}^{\mu/2T_{\rho}} d u \frac{\tanh{u}}{u}
\qquad (\gamma = 1.781),
                 \label{temperature.rho}
\end{equation}
\begin{equation}
T_{\rho} \ln{\left[1 +
\exp{ \left( \frac{\mu}{T_{\rho}} \right)}\right]} = \epsilon_{F},
                   \label{number.temperature.rho}
\end{equation}
where $\epsilon_{F} = \pi n_{f} d/m_{\perp}$ is the Fermi energy
\cite{foot1}.
Note once more that these equations coincide with the system which determines
the mean-field temperature $T_{c}^{(2D) MF}$ and $\mu(T_{c}^{(2D) MF})$
(see \cite{GGL,LSh.review}). This is evidently related to the mean-field
approximation used here and the limit $m_{z} \to \infty$.

Certainly in the simplest Landau theory one appears to have a phase
transition since $\rho$ takes on a non-zero value only below $T_\rho$. In
fact the parameter $\rho$ describes a gap only in the spectrum of the
{\it neutral} fermion field $\chi_\sigma(x)$. In describing the charged
(physical) fermion field $\psi_\sigma(x)$, $\rho$ only appears in conjunction
with $\theta$ in every correlation function. At zero temperature in closely
related four-Fermi models \cite{Witten,Silva} the pole in the neutral fermion
Green's function associated with the gap in the neutral fermion spectrum
is converted by the phase fluctuations into a branch cut in the Green's
function for charged fermions and we strongly suspect that similar behaviour
is present at finite temperature.

In the present approximation the neutral fermion Green's function ${\cal G}
(\mbox{\bf k},\omega)$ (\ref{neutg}) is identical to the BCS Green's function
but with energy gap $\rho$ rather than $\Delta$. As such the spectral density
of the neutral fermion Green's function, ${\cal G}$, is the sum of two delta
functions centered on the isolated poles of ${\cal G}$, $\omega = \pm
E(\mbox{\bf k})$ where $E(\mbox{\bf k}) = \sqrt{\xi^2(\mbox{\bf k}) + \rho^2}$
and one has zero density of states inside the gap $\rho$ \cite{Schrieffer}. In
the case of a branch cut (as postulated for the charged fermion
Green's function) the spectral density is non-zero for $\omega$ in the entire
range $-E(\mbox{\bf k}) \le \omega  \le E(\mbox{\bf k})$. Thus the spectral
density is smeared and one expects a corresponding smearing of the gap. An
illustrative example where a branch cut does lead to pseudogap behaviour is
given in \cite{Tchernyshyov}.

We believe that (see also discussion in \cite{Gusynin}) the expected smearing
of the  neutral fermion gap in the present model may well describe
the observed pseudogap behaviour. Furthermore the  fluctuations in $\theta$
are expected to convert the sharp phase transition to a smooth crossover.

\section{Results and Conclusions}

          The analytical and numerical investigation of the systems
(\ref{BKT.equation}), (\ref{rho.bound}), (\ref{number.rho})  and
(\ref{temperature.rho}), (\ref{number.temperature.rho}) yield the following
results.

One can show that at large densities ($\epsilon_{F} \gg
|\varepsilon_{b}| \gg T$) when $\mu \simeq \epsilon_{F}$
\begin{equation}
J(\epsilon_F,T,\rho) = \frac{1}{\pi} \epsilon_F \left[ 1 - \frac{1}{2} \int_{-
\infty}^\infty dx \frac{1}{\cosh^2 \sqrt{x^2 + \frac{\ds \rho^2}{\ds 4 T^2}}}
\right]
\end{equation}
and
\begin{equation}
J_z(\epsilon_F,T,\rho) = \frac{1}{2 \pi}
\frac{m_\perp}{m_z} w \left[ 1 - \frac{1}{2} \int_{-
\infty}^\infty dx \frac{1}{\cosh^2 \sqrt{x^2 + \frac{\ds \rho^2}{\ds 4 T^2}}}
\right]
\end{equation}
which yields
\begin{equation}
\alpha(\epsilon_{F}) = \frac{J_z}{J} = \frac{m_{\perp}}{m_{z}} \frac{w}
{2 \epsilon_{F}}
\label{anisotropy}
\end{equation}
which is independent of both $\rho$ and $T$. Thus even for a modest
anisotropy in the initial fermion Hamiltonian
one expects to obtain a large anisotropy (or equivalently a low value of
$\alpha$) in the effective quasi-2D XY model Hamiltonian. Indeed at all
densities we find small values for $\alpha$. Surprisingly $\alpha$ even
decreases as the carrier density increases. For this reason one can indeed
use the equation (\ref{BKT.equation}) to determine $T_{\rm BKT}^q$.
Therefore one expects the results obtained for the pure 2D model to persist to
the quasi-2D case.

Indeed for modest carrier densities ($\epsilon_F \ltwid 10^2 |\epsilon_b|$ or
the underdoped case) the PP (see Fig.~1) in the present model is roughly equal
in size to the superconducting region i.e.
$(T_\rho-T_{\rm BKT}^q)/T_{\rm BKT}^q \sim 1$. This allows us to
believe that in spite of the oversimplified character of the model proposed
in \cite{Gusynin} this approach may explain some of the observed normal state
anomalies of HTSC. For these densities we argue that the temperature $T_c$
for true LRO is probably roughly equal to $T_{\rm BKT}^q$. For very low
densities we find a large region between $T_c$ and $T_{\rm BKT}^q$ where the
system is  superconducting but one has two-dimensional and not
three-dimensional order. It is difficult to say whether such behaviour can be
observed experimentally because the densities are so low that the Fermi
surface is absent \cite{LSh.review,Turkowski}.

The model considered here is also a simple one. Certainly at sufficiently
large densities the BKT and PP region will disappear due to a direct
transition to the state with long-range order particularly for low anisotropy.
For example, in the case of an indirect
interaction in 2D, it has already been shown \cite{Turkowski} that the PP
region only exists at low carrier density.

The other results obtained in \cite{Gusynin} remain valid. In particular
one obtains a linear dependence of the critical superconducting temperature
($T_{\rm BKT}^q$) on the carrier density over a wide range of densities as is
seen in experiment.

For instance the kink in $\mu$ at $T = T_{\rho}$, (discussed
in reference \cite{Marel2}) occurs at the NP-PP boundary  or before
superconductivity appears. The ratio $2 \Delta/T_{\rm BKT}^q$ is also always
greater than the standard BCS value \cite{Abrikosov}.
The concentration behaviour of this ratio is consistent with experiment,
namely it decreases with increasing $\epsilon_{F}$.

It is also interesting to note that the qualitative difference between
the temperature dependence of the spin and charge correlations in the
normal state of the 2D attractive Hubbard model which was found in
\cite{Trivedi} acquires a natural explanation in the framework of the
approach used here. Indeed, while the neutral (with spin, but chargeless)
fermions have the gap $\rho$ in the PP, this gap should be smeared out
for the initial charged fermions due to the phase fluctuations.

To summarize, the presence of interlayer tunneling does not destroy the
pseudogap phase above the critical superconducting temperature since the
effective anisotropy (see (\ref{anisotropy})) is far larger than the simple
estimate of $m_z/m_{\perp}$.

\section*{Acknowledgments}
We gratefully acknowledge fruitful discussions with N.J.~Davidson,
V.P.~Gusynin, V.M.~Loktev and O.~Tchernyshyov. One of us (S.G.Sh) is
grateful to the members of the Department of Physics of the University of
Pretoria for hospitality. We acknowledge the financial support of
the Foundation for Research Development, Pretoria.

\begin{figure}
\setlength{\unitlength}{0.240900pt}
\ifx\plotpoint\undefined\newsavebox{\plotpoint}\fi
\sbox{\plotpoint}{\rule[-0.200pt]{0.400pt}{0.400pt}}%
\special{em:linewidth 0.4pt}%
\begin{picture}(1500,900)(0,0)
\font\gnuplot=cmr10 at 10pt
\gnuplot
\put(220,113){\special{em:moveto}}
\put(1436,113){\special{em:lineto}}
\put(220,113){\special{em:moveto}}
\put(220,877){\special{em:lineto}}
\put(220,113){\special{em:moveto}}
\put(240,113){\special{em:lineto}}
\put(1436,113){\special{em:moveto}}
\put(1416,113){\special{em:lineto}}
\put(198,113){\makebox(0,0)[r]{0}}
\put(220,495){\special{em:moveto}}
\put(240,495){\special{em:lineto}}
\put(1436,495){\special{em:moveto}}
\put(1416,495){\special{em:lineto}}
\put(198,495){\makebox(0,0)[r]{2}}
\put(220,877){\special{em:moveto}}
\put(240,877){\special{em:lineto}}
\put(1436,877){\special{em:moveto}}
\put(1416,877){\special{em:lineto}}
\put(198,877){\makebox(0,0)[r]{4}}
\put(220,113){\special{em:moveto}}
\put(220,133){\special{em:lineto}}
\put(220,877){\special{em:moveto}}
\put(220,857){\special{em:lineto}}
\put(220,68){\makebox(0,0){0}}
\put(463,113){\special{em:moveto}}
\put(463,133){\special{em:lineto}}
\put(463,877){\special{em:moveto}}
\put(463,857){\special{em:lineto}}
\put(463,68){\makebox(0,0){5}}
\put(706,113){\special{em:moveto}}
\put(706,133){\special{em:lineto}}
\put(706,877){\special{em:moveto}}
\put(706,857){\special{em:lineto}}
\put(706,68){\makebox(0,0){10}}
\put(950,113){\special{em:moveto}}
\put(950,133){\special{em:lineto}}
\put(950,877){\special{em:moveto}}
\put(950,857){\special{em:lineto}}
\put(950,68){\makebox(0,0){15}}
\put(1193,113){\special{em:moveto}}
\put(1193,133){\special{em:lineto}}
\put(1193,877){\special{em:moveto}}
\put(1193,857){\special{em:lineto}}
\put(1193,68){\makebox(0,0){20}}
\put(1436,113){\special{em:moveto}}
\put(1436,133){\special{em:lineto}}
\put(1436,877){\special{em:moveto}}
\put(1436,857){\special{em:lineto}}
\put(1436,68){\makebox(0,0){25}}
\put(220,113){\special{em:moveto}}
\put(1436,113){\special{em:lineto}}
\put(1436,877){\special{em:lineto}}
\put(220,877){\special{em:lineto}}
\put(220,113){\special{em:lineto}}
\put(45,495){\makebox(0,0){$T_c/|\varepsilon_{b}|$}}
\put(828,23){\makebox(0,0){$\epsilon_{F}/|\varepsilon_{b}|$}}
\put(804,476){\makebox(0,0)[l]{$T_{\rm BKT}^{\mbox{quasi}}$}}
\put(706,686){\makebox(0,0)[l]{$T_{\rho}$}}
\put(950,380){\makebox(0,0)[l]{$T_{\rm BKT}$}}
\put(1193,495){\makebox(0,0)[l]{${\bf BKT}$}}
\put(1193,686){\makebox(0,0)[l]{${\bf PP}$}}
\put(950,801){\makebox(0,0)[l]{${\bf NP}$}}
\sbox{\plotpoint}{\rule[-0.600pt]{1.200pt}{1.200pt}}%
\special{em:linewidth 1.2pt}%
\put(220,113){\special{em:moveto}}
\put(232,203){\special{em:lineto}}
\put(244,233){\special{em:lineto}}
\put(256,255){\special{em:lineto}}
\put(269,275){\special{em:lineto}}
\put(281,292){\special{em:lineto}}
\put(293,307){\special{em:lineto}}
\put(305,322){\special{em:lineto}}
\put(317,335){\special{em:lineto}}
\put(329,348){\special{em:lineto}}
\put(342,360){\special{em:lineto}}
\put(354,371){\special{em:lineto}}
\put(366,382){\special{em:lineto}}
\put(378,393){\special{em:lineto}}
\put(390,403){\special{em:lineto}}
\put(402,413){\special{em:lineto}}
\put(415,422){\special{em:lineto}}
\put(427,431){\special{em:lineto}}
\put(439,440){\special{em:lineto}}
\put(451,449){\special{em:lineto}}
\put(463,458){\special{em:lineto}}
\put(475,466){\special{em:lineto}}
\put(488,474){\special{em:lineto}}
\put(500,482){\special{em:lineto}}
\put(512,490){\special{em:lineto}}
\put(524,497){\special{em:lineto}}
\put(536,505){\special{em:lineto}}
\put(548,512){\special{em:lineto}}
\put(560,520){\special{em:lineto}}
\put(573,527){\special{em:lineto}}
\put(585,534){\special{em:lineto}}
\put(597,540){\special{em:lineto}}
\put(609,547){\special{em:lineto}}
\put(621,554){\special{em:lineto}}
\put(633,561){\special{em:lineto}}
\put(646,567){\special{em:lineto}}
\put(658,573){\special{em:lineto}}
\put(670,580){\special{em:lineto}}
\put(682,586){\special{em:lineto}}
\put(694,592){\special{em:lineto}}
\put(706,598){\special{em:lineto}}
\put(719,604){\special{em:lineto}}
\put(731,610){\special{em:lineto}}
\put(743,616){\special{em:lineto}}
\put(755,622){\special{em:lineto}}
\put(767,627){\special{em:lineto}}
\put(779,633){\special{em:lineto}}
\put(792,638){\special{em:lineto}}
\put(804,644){\special{em:lineto}}
\put(816,650){\special{em:lineto}}
\put(828,655){\special{em:lineto}}
\put(840,660){\special{em:lineto}}
\put(852,666){\special{em:lineto}}
\put(864,671){\special{em:lineto}}
\put(877,676){\special{em:lineto}}
\put(889,681){\special{em:lineto}}
\put(901,686){\special{em:lineto}}
\put(913,692){\special{em:lineto}}
\put(925,697){\special{em:lineto}}
\put(937,701){\special{em:lineto}}
\put(950,706){\special{em:lineto}}
\put(962,711){\special{em:lineto}}
\put(974,716){\special{em:lineto}}
\put(986,721){\special{em:lineto}}
\put(998,726){\special{em:lineto}}
\put(1010,731){\special{em:lineto}}
\put(1023,735){\special{em:lineto}}
\put(1035,740){\special{em:lineto}}
\put(1047,745){\special{em:lineto}}
\put(1059,749){\special{em:lineto}}
\put(1071,754){\special{em:lineto}}
\put(1083,758){\special{em:lineto}}
\put(1096,763){\special{em:lineto}}
\put(1108,767){\special{em:lineto}}
\put(1120,772){\special{em:lineto}}
\put(1132,776){\special{em:lineto}}
\put(1144,781){\special{em:lineto}}
\put(1156,785){\special{em:lineto}}
\put(1168,789){\special{em:lineto}}
\put(1181,794){\special{em:lineto}}
\put(1193,798){\special{em:lineto}}
\put(1205,802){\special{em:lineto}}
\put(1217,807){\special{em:lineto}}
\put(1229,811){\special{em:lineto}}
\put(1241,815){\special{em:lineto}}
\put(1254,819){\special{em:lineto}}
\put(1266,823){\special{em:lineto}}
\put(1278,827){\special{em:lineto}}
\put(1290,831){\special{em:lineto}}
\put(1302,835){\special{em:lineto}}
\put(1314,840){\special{em:lineto}}
\put(1327,844){\special{em:lineto}}
\put(1339,848){\special{em:lineto}}
\put(1351,851){\special{em:lineto}}
\put(1363,855){\special{em:lineto}}
\put(1375,859){\special{em:lineto}}
\put(1387,863){\special{em:lineto}}
\put(1400,867){\special{em:lineto}}
\put(1412,871){\special{em:lineto}}
\put(1424,875){\special{em:lineto}}
\put(1431,877){\special{em:lineto}}
\sbox{\plotpoint}{\rule[-0.500pt]{1.000pt}{1.000pt}}%
\special{em:linewidth 1.0pt}%
\put(220,113){\usebox{\plotpoint}}
\put(220.00,113.00){\usebox{\plotpoint}}
\put(238.56,122.28){\usebox{\plotpoint}}
\multiput(244,125)(18.564,9.282){0}{\usebox{\plotpoint}}
\put(257.15,131.53){\usebox{\plotpoint}}
\put(275.89,140.44){\usebox{\plotpoint}}
\multiput(281,143)(18.564,9.282){0}{\usebox{\plotpoint}}
\put(294.45,149.73){\usebox{\plotpoint}}
\put(313.02,159.01){\usebox{\plotpoint}}
\multiput(317,161)(18.564,9.282){0}{\usebox{\plotpoint}}
\put(331.69,168.04){\usebox{\plotpoint}}
\put(350.69,176.34){\usebox{\plotpoint}}
\multiput(354,178)(18.564,9.282){0}{\usebox{\plotpoint}}
\put(369.25,185.63){\usebox{\plotpoint}}
\put(387.81,194.91){\usebox{\plotpoint}}
\multiput(390,196)(19.159,7.983){0}{\usebox{\plotpoint}}
\put(406.82,203.23){\usebox{\plotpoint}}
\put(425.85,211.52){\usebox{\plotpoint}}
\multiput(427,212)(18.564,9.282){0}{\usebox{\plotpoint}}
\put(444.62,220.34){\usebox{\plotpoint}}
\multiput(451,223)(18.564,9.282){0}{\usebox{\plotpoint}}
\put(463.39,229.16){\usebox{\plotpoint}}
\put(482.64,236.94){\usebox{\plotpoint}}
\multiput(488,239)(18.564,9.282){0}{\usebox{\plotpoint}}
\put(501.47,245.61){\usebox{\plotpoint}}
\put(520.63,253.60){\usebox{\plotpoint}}
\multiput(524,255)(19.159,7.983){0}{\usebox{\plotpoint}}
\put(539.79,261.58){\usebox{\plotpoint}}
\put(558.95,269.56){\usebox{\plotpoint}}
\multiput(560,270)(19.372,7.451){0}{\usebox{\plotpoint}}
\put(578.25,277.19){\usebox{\plotpoint}}
\multiput(585,280)(19.159,7.983){0}{\usebox{\plotpoint}}
\put(597.41,285.17){\usebox{\plotpoint}}
\put(616.57,293.15){\usebox{\plotpoint}}
\multiput(621,295)(19.690,6.563){0}{\usebox{\plotpoint}}
\put(636.08,300.19){\usebox{\plotpoint}}
\put(655.35,307.90){\usebox{\plotpoint}}
\multiput(658,309)(19.159,7.983){0}{\usebox{\plotpoint}}
\put(674.64,315.55){\usebox{\plotpoint}}
\put(693.99,323.00){\usebox{\plotpoint}}
\multiput(694,323)(19.690,6.563){0}{\usebox{\plotpoint}}
\put(713.56,329.91){\usebox{\plotpoint}}
\multiput(719,332)(19.690,6.563){0}{\usebox{\plotpoint}}
\put(733.10,336.88){\usebox{\plotpoint}}
\put(752.52,344.17){\usebox{\plotpoint}}
\multiput(755,345)(19.159,7.983){0}{\usebox{\plotpoint}}
\put(771.88,351.63){\usebox{\plotpoint}}
\put(791.66,357.90){\usebox{\plotpoint}}
\multiput(792,358)(19.690,6.563){0}{\usebox{\plotpoint}}
\put(811.16,364.98){\usebox{\plotpoint}}
\multiput(816,367)(19.690,6.563){0}{\usebox{\plotpoint}}
\put(830.71,371.90){\usebox{\plotpoint}}
\put(850.40,378.47){\usebox{\plotpoint}}
\multiput(852,379)(19.159,7.983){0}{\usebox{\plotpoint}}
\put(869.80,385.79){\usebox{\plotpoint}}
\multiput(877,388)(19.690,6.563){0}{\usebox{\plotpoint}}
\put(889.55,392.18){\usebox{\plotpoint}}
\put(909.24,398.75){\usebox{\plotpoint}}
\multiput(913,400)(19.690,6.563){0}{\usebox{\plotpoint}}
\put(928.93,405.31){\usebox{\plotpoint}}
\put(948.70,411.60){\usebox{\plotpoint}}
\multiput(950,412)(19.690,6.563){0}{\usebox{\plotpoint}}
\put(968.40,418.13){\usebox{\plotpoint}}
\multiput(974,420)(19.690,6.563){0}{\usebox{\plotpoint}}
\put(988.09,424.70){\usebox{\plotpoint}}
\put(1008.01,430.50){\usebox{\plotpoint}}
\multiput(1010,431)(19.838,6.104){0}{\usebox{\plotpoint}}
\put(1027.84,436.61){\usebox{\plotpoint}}
\multiput(1035,439)(19.690,6.563){0}{\usebox{\plotpoint}}
\put(1047.53,443.18){\usebox{\plotpoint}}
\put(1067.40,449.10){\usebox{\plotpoint}}
\multiput(1071,450)(19.690,6.563){0}{\usebox{\plotpoint}}
\put(1087.21,455.29){\usebox{\plotpoint}}
\put(1106.96,461.65){\usebox{\plotpoint}}
\multiput(1108,462)(20.136,5.034){0}{\usebox{\plotpoint}}
\put(1126.92,467.31){\usebox{\plotpoint}}
\multiput(1132,469)(19.690,6.563){0}{\usebox{\plotpoint}}
\put(1146.67,473.67){\usebox{\plotpoint}}
\put(1166.56,479.52){\usebox{\plotpoint}}
\multiput(1168,480)(19.838,6.104){0}{\usebox{\plotpoint}}
\put(1186.47,485.37){\usebox{\plotpoint}}
\multiput(1193,487)(19.690,6.563){0}{\usebox{\plotpoint}}
\put(1206.34,491.33){\usebox{\plotpoint}}
\put(1226.26,497.09){\usebox{\plotpoint}}
\multiput(1229,498)(20.136,5.034){0}{\usebox{\plotpoint}}
\put(1246.26,502.62){\usebox{\plotpoint}}
\multiput(1254,505)(20.136,5.034){0}{\usebox{\plotpoint}}
\put(1266.27,508.09){\usebox{\plotpoint}}
\put(1286.14,514.04){\usebox{\plotpoint}}
\multiput(1290,515)(19.690,6.563){0}{\usebox{\plotpoint}}
\put(1306.00,520.00){\usebox{\plotpoint}}
\put(1325.96,525.68){\usebox{\plotpoint}}
\multiput(1327,526)(20.136,5.034){0}{\usebox{\plotpoint}}
\put(1346.08,530.77){\usebox{\plotpoint}}
\multiput(1351,532)(19.690,6.563){0}{\usebox{\plotpoint}}
\put(1365.95,536.74){\usebox{\plotpoint}}
\put(1386.08,541.77){\usebox{\plotpoint}}
\multiput(1387,542)(19.838,6.104){0}{\usebox{\plotpoint}}
\put(1406.02,547.51){\usebox{\plotpoint}}
\multiput(1412,549)(20.136,5.034){0}{\usebox{\plotpoint}}
\put(1426.11,552.70){\usebox{\plotpoint}}
\put(1436,556){\usebox{\plotpoint}}
\sbox{\plotpoint}{\rule[-0.600pt]{1.200pt}{1.200pt}}%
\special{em:linewidth 1.2pt}%
\put(232,122){\special{em:moveto}}
\put(244,130){\special{em:lineto}}
\put(256,138){\special{em:lineto}}
\put(269,146){\special{em:lineto}}
\put(281,154){\special{em:lineto}}
\put(293,162){\special{em:lineto}}
\put(305,169){\special{em:lineto}}
\put(317,176){\special{em:lineto}}
\put(329,184){\special{em:lineto}}
\put(342,191){\special{em:lineto}}
\put(354,198){\special{em:lineto}}
\put(366,205){\special{em:lineto}}
\put(378,211){\special{em:lineto}}
\put(390,218){\special{em:lineto}}
\put(402,224){\special{em:lineto}}
\put(415,231){\special{em:lineto}}
\put(427,237){\special{em:lineto}}
\put(439,243){\special{em:lineto}}
\put(451,249){\special{em:lineto}}
\put(463,255){\special{em:lineto}}
\put(475,261){\special{em:lineto}}
\put(488,267){\special{em:lineto}}
\put(500,273){\special{em:lineto}}
\put(512,278){\special{em:lineto}}
\put(524,284){\special{em:lineto}}
\put(536,290){\special{em:lineto}}
\put(548,295){\special{em:lineto}}
\put(560,300){\special{em:lineto}}
\put(573,306){\special{em:lineto}}
\put(585,311){\special{em:lineto}}
\put(597,316){\special{em:lineto}}
\put(609,321){\special{em:lineto}}
\put(621,327){\special{em:lineto}}
\put(633,332){\special{em:lineto}}
\put(646,336){\special{em:lineto}}
\put(658,341){\special{em:lineto}}
\put(670,346){\special{em:lineto}}
\put(682,351){\special{em:lineto}}
\put(694,356){\special{em:lineto}}
\put(706,361){\special{em:lineto}}
\put(719,366){\special{em:lineto}}
\put(731,370){\special{em:lineto}}
\put(743,375){\special{em:lineto}}
\put(755,380){\special{em:lineto}}
\put(767,384){\special{em:lineto}}
\put(779,389){\special{em:lineto}}
\put(792,393){\special{em:lineto}}
\put(804,398){\special{em:lineto}}
\put(816,402){\special{em:lineto}}
\put(828,407){\special{em:lineto}}
\put(840,411){\special{em:lineto}}
\put(852,415){\special{em:lineto}}
\put(864,420){\special{em:lineto}}
\put(877,424){\special{em:lineto}}
\put(889,428){\special{em:lineto}}
\put(901,432){\special{em:lineto}}
\put(913,437){\special{em:lineto}}
\put(925,441){\special{em:lineto}}
\put(937,445){\special{em:lineto}}
\put(950,449){\special{em:lineto}}
\put(962,453){\special{em:lineto}}
\put(974,457){\special{em:lineto}}
\put(986,461){\special{em:lineto}}
\put(998,465){\special{em:lineto}}
\put(1010,469){\special{em:lineto}}
\put(1023,473){\special{em:lineto}}
\put(1035,477){\special{em:lineto}}
\put(1047,481){\special{em:lineto}}
\put(1059,485){\special{em:lineto}}
\put(1071,489){\special{em:lineto}}
\put(1083,493){\special{em:lineto}}
\put(1096,496){\special{em:lineto}}
\put(1108,500){\special{em:lineto}}
\put(1120,504){\special{em:lineto}}
\put(1132,508){\special{em:lineto}}
\put(1144,511){\special{em:lineto}}
\put(1156,515){\special{em:lineto}}
\put(1168,519){\special{em:lineto}}
\put(1181,523){\special{em:lineto}}
\put(1193,526){\special{em:lineto}}
\put(1205,530){\special{em:lineto}}
\put(1217,534){\special{em:lineto}}
\put(1229,537){\special{em:lineto}}
\put(1241,541){\special{em:lineto}}
\put(1254,544){\special{em:lineto}}
\put(1266,548){\special{em:lineto}}
\put(1278,551){\special{em:lineto}}
\put(1290,555){\special{em:lineto}}
\put(1302,558){\special{em:lineto}}
\put(1314,562){\special{em:lineto}}
\put(1327,565){\special{em:lineto}}
\put(1339,569){\special{em:lineto}}
\put(1351,572){\special{em:lineto}}
\put(1363,576){\special{em:lineto}}
\put(1375,579){\special{em:lineto}}
\put(1387,583){\special{em:lineto}}
\put(1400,586){\special{em:lineto}}
\put(1412,589){\special{em:lineto}}
\put(1424,593){\special{em:lineto}}
\put(1436,596){\special{em:lineto}}
\end{picture}

\vspace{1cm}
\caption{
$T_{\rm BKT}^q$, $T_{\rm BKT}$ (dots) and $T_{\rho}$ versus the
noninteracting fermion density.  The regions of the normal phase (NP),
pseudogap phase (PP) and BKT phase are indicated.
Note that in the approximation used the value of $T_{\rho}$ is the
same for the 2D and quasi-2D models. We assumed that
$m_{z}/m_{\perp} = 100$ and $(m_{z}d^{2} |\varepsilon_{b}|)^{-1}=0.1$.}
\end{figure}
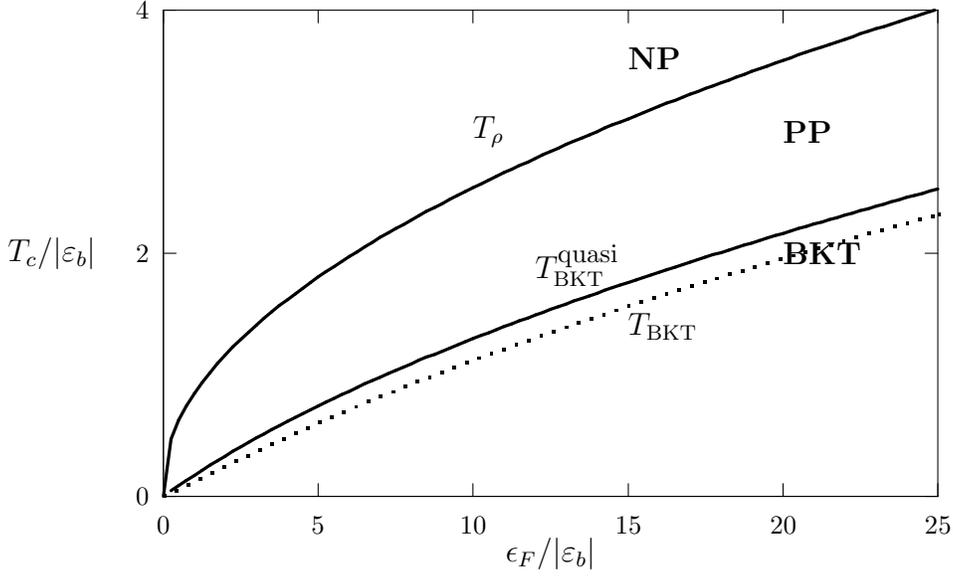

\begin{thebibliography}{99}
\bibitem{foot1} The Fermi energy for the quasi-2D system
$\epsilon_{F} = \epsilon_{F}^{2D} d$, where
$\epsilon_{F}^{2D} = \pi n_{f}/m_\perp$
is the Fermi energy for the 2D system with the quadratic dispersion
law.
\end{thebibliography}
\end{document}